\documentclass[reprint, aps, amsmath,amssymb]{revtex4-1}

\usepackage{graphicx}
\usepackage{dcolumn}
\usepackage{bm}

\newcommand{\varA}[1]{{\operatorname{#1}}}

\begin{document}

\preprint{APS/123-QED}

\title{Simplified Threshold Phenomena in Hypo- and Hypercoagulation}

\author{Jayavel Arumugam}
\affiliation{%
 jayavel.arumugam.17@gmail.com, Department of Mechanical Engineering, Texas A\&M University
}%





\begin{abstract}
We discuss two threshold phenomena in blood coagulation dynamics using a simplified model. This perspective of the underlying complex phenomena is expected to aid the understanding and characterization of many blood coagulation pathologies with altered protein dynamics. 

\end{abstract}

\pacs{Valid PACS appear here}
\maketitle


\section{The Unusual Suspects}


It is usual to think of nonlinear dynamics while modeling problems with the heart \cite{guevara1981phase} and the brain \cite{fitzhugh1955mathematical} cells. Some known hematological disorders that are of dynamical nature are centered around blood cells \cite{mackey1987dynamical}. 

In this letter, we try to bring to light many more dynamical problems with blood coagulation that are centered around proteins and, in particular, an usual suspect named thrombin. We do this by using a simplified dynamical model for an important part of the blood coagulation cascade concerning thrombin. We focus on two of its essential threshold behavior. 

\section{Complexity and Clutter} 







%
Many pathologies of blood coagulation are understood in terms of altered states of blood chemistry \cite{rapaport1993blood}. Decades of experimental work studying chemical kinetics have led to models for atomic parts of the system built using individual reactions. Such models describing kinetics using $M$ reactions are usually concatenated to the form \cite{hockin2002model},
\begin{align} 
\frac{d [Y_{i}]}{d t} 
&= \sum_{j=1}^{M} S_{ij}
r_{j}([\mathbf{Y}]; \mathbf{k}_j ), \\
s.t. \quad  
[Y_{i}](0) 
& = [Y_{i}]^{0}, \notag \\
\mathbf{S}[\mathbf{Y}] &= \mathbf{S}[\mathbf{Y}_{}]^{0}, \notag \\
[Y_{i}] &\geq 0, \notag
\end{align}
\noindent where $S_{ij}$ are stoichiometric coefficients (with $i = 1, ... ,N, \quad j = 1, ..., M$), $r_{j}$ are reaction rates,  $\mathbf{k}_j$ are parameters of the reaction rates, and $[Y_{i}]^{0}$ are the initial plasma composition. However, pathologies involving alterations in dynamics of the blood chemical reactions have not been well understood. 

\begin{figure}[htbp]
\centering
\includegraphics[width=0.5 \textwidth]{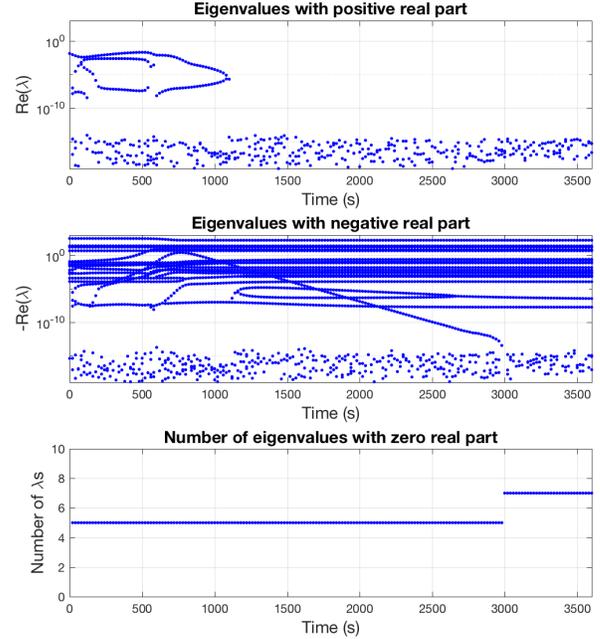}
\caption[Eigenvalues of the Jacobian of the reaction rates]{Eigenvalues of the Jacobian of reaction rates as a function of time. The model, which describes dynamics of 34 proteins using 42 reactions, is very stiff and the solution trajectories are unstable in many directions. The zero eigenvalues are due to 9 stoichiometric constraints. Only 5 out of the 9 constraints are met due to numerical precision.} 
\label{fig:EigenvaluesThrombin}
\end{figure}

Concentration of proteins involved in coagulation and rates of reactions vary by orders of magnitude. Essentially, picomoles of trigger results in the formation of hundreds of nanomoles of certain enzymes. This in turn results in macroscopic formation of polymeric-fibrous clots. The system has so far been impenetrable towards understanding the of the sum of the parts. Partly, this could reasoned based on Figure \ref{fig:EigenvaluesThrombin}. Eigenvalues (plotted in log scale) of the Jacobian of the reaction rates of clotting obtained from a specific model \cite{hockin2002model} portrays a ghastly serpentine picture. 

Geometrical analysis of chemical kinetic systems such as in combustion show such complexity \cite{ren2006geometry}. Possible coexistence of qualitatively different local instabilities along a trajectory in such multi-dimensional systems \cite{okushima2005finite} obfuscate the process. Needless to say, the formidability of formulating meaningful hypothesis and designing good experiments have lead to ambiguous theories \cite{hemker2012there}, experimental data that is often hard to interpret, and sometimes even compromised \cite{knappe2015application}. Nevertheless, artful analysis of such models have affirmed them to be useful thinking tools \cite{mann2012there} and certain aspects of the dynamics show sustained promise in a clinical setting \cite{brummel2014modeling}. 

Recently, there have been attempts to use black box machine learning tools to practically apply the system \cite{arumugam2016random}. But the urge to understand the nature of things have further led to a simplified model for the essential dynamics \cite{jayavel2017dissertation} (also \cite{jayavel2017arxivsimplemodel}).  Here we attempt to draw it. 


\section{Simplification and Structure}
We consider the essential dynamics of concentration just four species: i) bursting enzyme thrombin $\varA{[IIa]}$, ii) precursor prothrombin $\varA{[II]}$, iii) inhibitor antithrombin III $\varA{[ATIII]}$, and iv) by-product thrombin-antithrombin $\varA{[IIa-ATIII]}$. 

\subsection{Simplified Model} 
The simplified model is given by, 

\begin{align}
&\frac{d}{dt} \varA{[II]} = - K_{S} - K_{P}  \varA{[II]} \varA{[IIa]}   \notag \\
&\frac{d}{dt} \varA{[IIa]} = K_{S} + K_{P}  \varA{[II]} \varA{[IIa]}   -  K_{I}  \varA{[IIa]}  \varA{[ATIII]} \notag \\
&\frac{d}{dt} \varA{[AT]} = - K_{I}  \varA{[IIa]}  \varA{[ATIII]} \notag 
\label{eq:simplfiedModelReactionsRates}
\end{align}

\noindent The rate of change of the by-product is given by $\frac{d}{dt} \varA{[IIa-ATIII]} = K_{I}  \varA{[IIa]}  \varA{[ATIII]}$ so that stoichiometry is preserved. The `macroscopic' model phenomenologically accounts for the effect of myriads of other `microscopic' species in the rate parameters.  Further, this drastic simplification deals with the intrinsic complexity by means of switching, i.e., the rate parameters abruptly switch based on the amount of enzyme $\varA{[IIa]}$ present (Table \ref{tab:switchingConditions}). Moreover, the switching was designed to reflect the two threshold phenomena discussed next. 

\begin{table}[htbp]
\caption{Switching Condition. Estimates for the physiological mean plasma are $k_{s} = 0.005$ nMs$^{-1}$ (for 5 pM of trigger), $k_{i2} = 1.9551E-07$ nM$^{-1}$s$^{-1}$, $k_{i1} = 6.347E-06$ nM$^{-1}$s$^{-1}$, and $k_{p} = 4.155E-05$ nM$^{-1}$s$^{-1}$. }
\label{tab:switchingConditions}       
\begin{tabular}{l|c|c}
\hline\noalign{\smallskip}
 & [IIa] $ < $ 2 nM & [IIa] $ \geq $ 2 nM  \\
\noalign{\smallskip}\hline\noalign{\smallskip}
$K_{S}$ & $k_{s} > 0$ & 0 \\
$K_{I}$ & $k_{i2} > 0$ & $k_{i1} > 0$ \\
$K_{P}$ & 0 & $k_{p} > 0$ \\
\noalign{\smallskip}\hline
\end{tabular}
\end{table}

\begin{figure}[htbp]
\centering
\includegraphics[width=0.47 \textwidth]{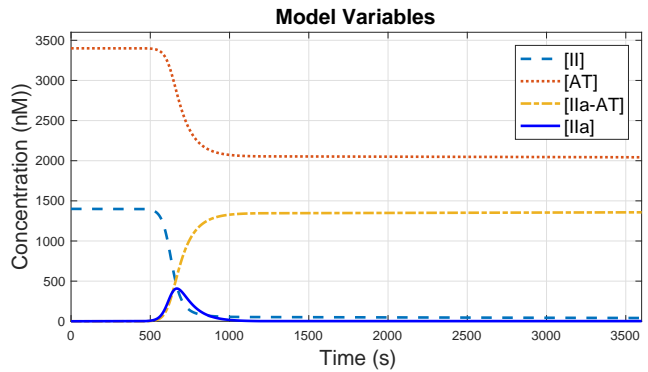}
\caption[]{Typical response of the model variables during clotting for the physiological mean: $\varA{[ATIII]^{0}} = 3400$ nM, $\varA{[II]^{0}} = 1400$ nM, $\varA{[IIa]^{0}} = \varA{[IIa-ATIII]^{0}} = 0$.} 
\label{fig:clotResponse}
\end{figure} 

There are essentially two threshold phenomena exhibited by the simplified model: i) all-or-none clot initiation and ii) sustained clot propagation. To avoid false triggers, clot initiates only if there are sufficient amounts of initiation triggers present; clot better not initiate due to weak triggers. Further, clotting better be put in check. There are, normally, more than enough inhibitors in the blood to do so. In certain abnormal cases when they are inadequate, the model predicts that clotting need not terminate. The two thresholds arise from the two necessary functional roles. The former concerns with hypocoagulation while the latter is a signature of hypercoagulation. 

We describe these behavior primarily using 2D projections of the phase portrait of the simplified model. In addition, we use log scales, as and when needed, in order to tame aspects of the ghastly serpent. To aid interpretation of the figures, we note that thrombin, i.e., IIa, is an enzyme that catalyzes clot formation and its presence signifies clotting in almost all cases. 

\subsection{To Clot or Not to Clot?}

\begin{figure}[htbp]
\centering
\includegraphics[width=0.5 \textwidth]{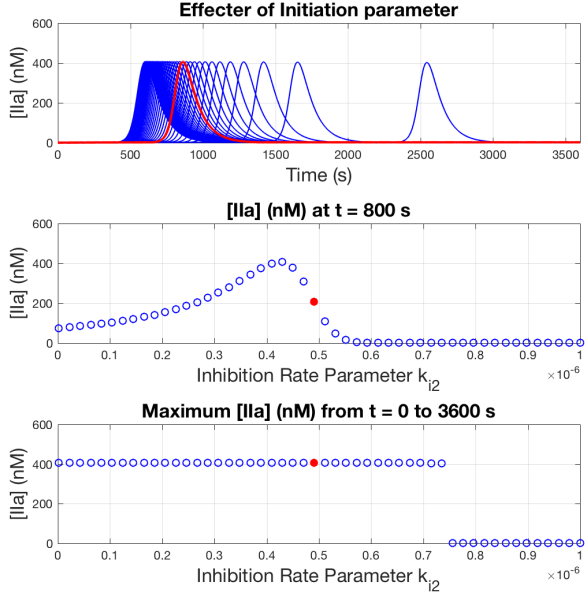}
\caption[]{Simplified all-or-none clot initiation. Data from a specific parameter value is highlighted in red.} 
\label{fig:initiationThreshold}
\end{figure}

\begin{figure}[htbp]
\centering
\includegraphics[width=0.45 \textwidth]{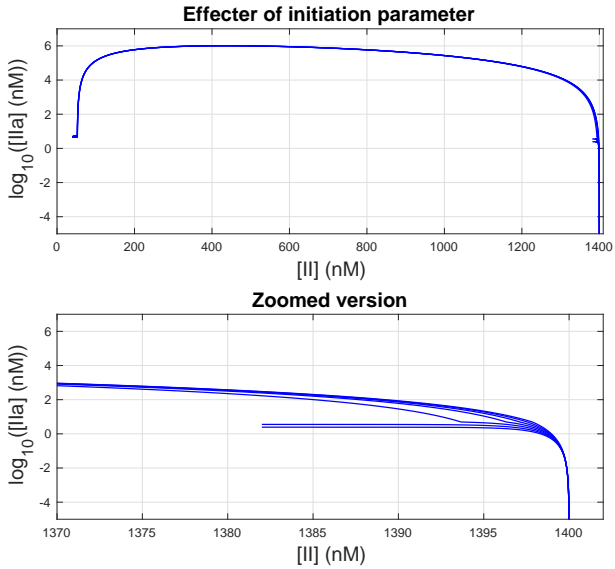}
\caption[2D projections of reaction trajectories depicting thrombin initiation.]{2D projections of reaction trajectories depicting thrombin initiation.} 
\label{fig:initiationphasePortrait}
\end{figure}

Figure \ref{fig:initiationThreshold} shows thrombin generation profiles for varying values of $k_{i2}$. $\varA{[IIa]}$ concentration at $t = 800 s$ show a delay in clot initiation with larger values of $k_{i2}$. Dependence on maximum values of $\varA{[IIa]}$ on $k_{i2}$ suggest clotting does not initiate in the time frame for values of $k_{i2}$ above a certain threshold. Figure \ref{fig:initiationphasePortrait} shows the projection of the trajectories on a 2D plane. The zoomed version better shows the threshold effect of the $k_{i2}$ on clot initiation. Hypocoagulation occur in many well-known forms of hemophilia when certain proteins are deficient \cite{rapaport1993blood}.

\subsection{To Terminate or To Propagate?}

\begin{figure}[tbp]
\centering
\includegraphics[width=0.5 \textwidth]{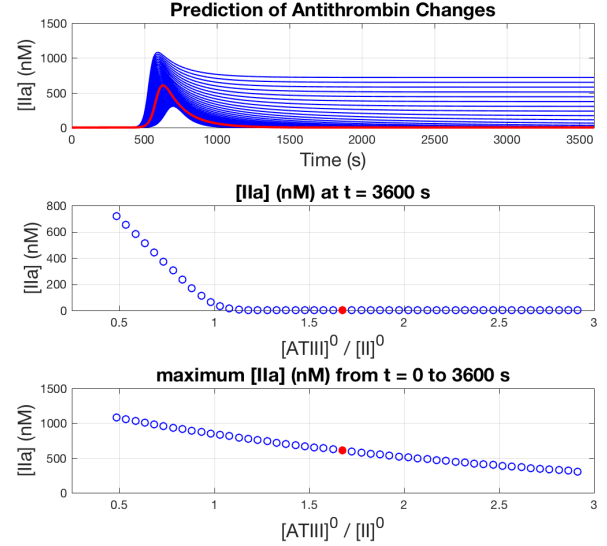}
\caption[]{Simplified threshold response of clot propagation based on antithrombin (and prothrombin ratio).} 
\label{fig:effectOfAntithrombin}
\end{figure}

\begin{figure}[tbp]
\centering
\includegraphics[width=0.45 \textwidth]{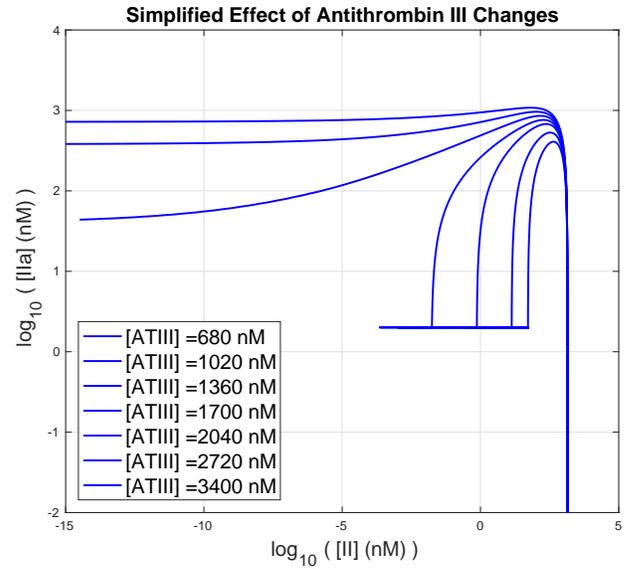}
\caption[2D projections of reaction trajectories due to Antithrombin III variation.]{2D projections of reaction trajectories due to Antithrombin III variation.} 
\label{fig:ATIIIphasePortrait}
\end{figure}

Figure \ref{fig:effectOfAntithrombin} shows the effect of initial $\varA{[ATIII]_{}^{0}}$ on $\varA{[IIa]}$ dynamics. Activity of $\varA{[IIa]}$ sustains for certain values of initial $\varA{[ATIII]_{}^{0}}$. This is seen in value of $\varA{[IIa]}$ at $t = 3600 s$. In this model, $\varA{[IIa]}$ runs out of $\varA{[ATIII]}$. Trajectory projections on a 2D plane in Figure \ref{fig:ATIIIphasePortrait} show the marked difference between normal thrombin trajectory and sustained activity. This fact, observed in experimental model systems \cite{allen2004impact}, could have escaped most experimentalists if they had measured $\varA{[IIa-ATIII]}$ to infer $\varA{[IIa]}$ or used a very common method that was recently found to be compromised \cite{knappe2015application} (and explainable due to sustained $\varA{[IIa]}$ activity). 

ATIII is deficient in many hypercoagulative pathologies \cite{thaler1981antithrombin}. This model predicts $\varA{[ATIII]^{0}}/\varA{[II]^{0}}  \geq 1 $  (for $\varA{[II]^{0}} = 1400$ nM, $\varA{[ATIII]^{0}} = 1400$ nM or 41\% of physiological mean \cite{hockin2002model}) to avoid hypercoagulation. Reported data on hypercoagulative disseminated intravascular coagulation (DIC) patients has a mean  of 66\% $\varA{[ATIII]^{0}}$ \cite{bick1980antithrombin}. This corroboration is hindered due to the fact that $\varA{[II]^{0}}$ values are not reported together with $\varA{[ATIII]^{0}}$ (very likely, the significance of the ratio is first highlighted by this study). Further, many ATIII assays measure relative percent of ATIII and there are two different reported physiological mean values \cite{hockin2002model,kalafatis1997regulation} (the other value is 2500 nM). 

ATIII deficiency could be due to faulty synthesis, increased consumption, loss (such as filtration in kidneys), etc. \cite{thaler1981antithrombin}. The proteins not explicitly accounted for in this model would precisely determine the threshold value of the ratio. Models accounting for other inhibitors like activated protein C could better explain the nature of stability in the different regions of the phase portrait. 

\section{Excitement and Enthusiasm}

\begin{figure}[tbp]
\centering
\includegraphics[width=0.4 \textwidth]{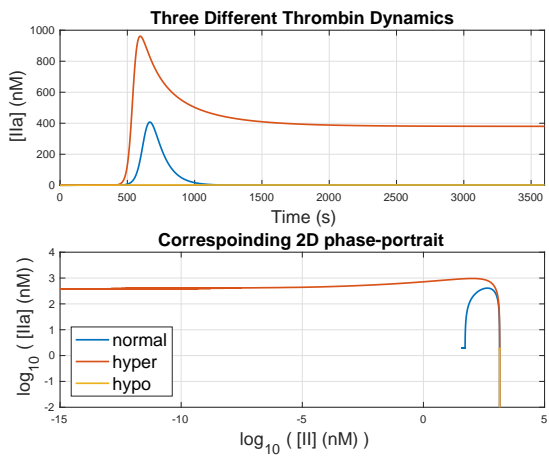}
\caption[]{Summary of threshold phenomena in the simplified model.} 
\label{fig:thresholdSummary}
\end{figure}

The three physiologically different simulation responses including the two threshold responses are summarized in Figure \ref{fig:thresholdSummary}. In the simplified model response, there is no clot initiation in hypocoagulation, and there is sustained $\varA{[IIa]}$ activity in hypercoagulation. The extent to which `normal' trajectories remain normal demands judicious experimental studies. For example, circulation could amplify seemingly minor differences in clotting at a particular location due to the effect of time delay. 

Circulation along with diffusion also introduce a spatial aspect to the reaction trajectories. Another key contribution of this simplified analysis is that it offers an explanation for the effect of clotting in one part of the body on clotting elsewhere using sustained thrombin activity. Such activity could potentially propagate downstream leading to dire consequences as in the case of DIC \cite{toh2003disseminated}. 
Further, understanding and modeling the feedback mechanisms and drug administration that control the amount of proteins in the blood such as II and ATIII, and the reaction dynamics \cite{hemker2016century}, opens up a way to understand the possibly dynamical \cite{mackey1987dynamical} nature of many blood coagulation pathologies. 

Promisingly, as the perspective suggests, measurements made somewhere in the body could reveal a lot on what is likely to happen at many other places. 



\section{Introduction}
We offered a simplified depiction of the blood coagulation cascade and emphasized three essentially different physiological responses. We hope this paper brings to light, to the nonlinear dynamics community, the similar nature of blood coagulation disorders associated with proteins and, in the process, eventually catalyzes advances in better diagnosis and treatment. Though the simplified picture is far from reality, it would be wronger than wrong to dismiss its implications. We also expect that the serpent we have portrayed is only a part of a larger elephant. Thrombin is  multi-functional and plays an active role in pathways associated with diseases such as inflammation and cancer. 


\bibliographystyle{aipauth4-1}       
\bibliography{bibdataBlood_03Bio052017}   

\end{document}